\documentclass[aps,pra,superscriptaddress,floatfix,showpacs,10pt,twocolumn]{revtex4}
\usepackage[dvips]{graphicx}
\usepackage{amssymb}
\usepackage{amsmath}
\begin{document}
\title{Concentration for unknown atomic entangled states via cavity decay}
\author{Zhuo-Liang Cao}
\email{zlcao@ahu.edu.cn(Corresponding~Author)} \affiliation{Schoolof
Physics \& Material Science, Anhui University, Hefei, 230039,
People's Republic of China}
\author{Li-Hua Zhang}
\email{zhanglh@aqtc.edu.cn} \affiliation{Schoolof Physics \&
Material Science, Anhui University, Hefei, 230039, People's Republic
of China} \affiliation{Department of Physics, Anqing Teacher
College, Anqing, 246011, People's Republic of China}
\author{Ming Yang}
\email{mingyang@ahu.edu.cn} \affiliation{Schoolof Physics \&
Material Science, Anhui University, Hefei, 230039, People's Republic
of China}

\begin{abstract}
We present a physical scheme for entanglement concentration of unknown
atomic entangled states via cavity decay. In the scheme, the atomic state is
used as stationary qubit and photonic state as flying qubit, and a close
maximally entangled state can be obtained from pairs of partially entangled
states probabilistically.
\end{abstract}

 \maketitle

Entanglement plays an important role ~\cite{Bennett1, Bennett2, Ekert,
Bennett3} in quantum information and computation. To fulfill perfect quantum
information processing (QIP) the quantum channel must be maximally entangled
usually. But, in the real processing, storage and transmission of quantum
states, the maximally entangled states usually collapse into nonmaximally
entangled ones or even mixed states because of noise. To achieve faithful
QIP, we must convert the nonmaximally entangled states into pure maximally
entangled ones. There are many theoretical and experimental schemes that can
achieve this conversion, such as entanglement concentration (pure
nonmaximally entangled states case)~\cite{Bennett4}, entanglement
purification or distillation (mixed entangled states case)~\cite{Bennett5}.

Recently, Zhao \textit{et al}.~\cite{Zhao} present a practical scheme for
entanglement concentration with the polarization beam splitters. For pure
nonmaximally photonic polarized entangled states, S. Bose \textit{et al}.~%
\cite{Bose} have proposed a scheme, to realize entanglement
purification via entanglement swapping. In the above two
proposals~\cite{Zhao, Bose}, they utilize photonic state as both
stationary qubits and flying qubits. Photonic qubits, despite its
perfection as flying qubits, is not ideal for stationary qubits.
Oppositely, the atomic states are known as the ideal stationary
qubits in quantum information theory. In the realm of QIP, cavity
QED is one of the promising candidates dealing with atomic qubits.
There are several entanglement concentration and purification
schemes for atomic entangled states in cavity QED~\cite{yang, cao,
ye}, where the cavity decay is neglected.

However, spontaneous and detected decay are unavoidable in practical quantum
information processing~\cite{plenio} in cavity QED system. So, teleportation
and generation of atomic entanglement is more general when taking cavity
decay into consideration~\cite{Bose2, zou, yu, Chimczak1, Chimczak2}.
Obviously, it is possible to generate an entangled state between two atoms,
located in two different decay cavities, respectively. Here, we present a
scheme for entanglement concentration of unknown atomic entangled states via
cavity decay.

\begin{figure}[tbp]
\includegraphics[scale=0.4, angle=90]{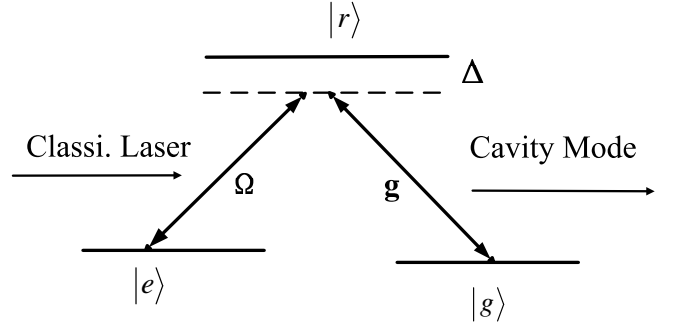}
\caption{Level structure of the atoms used in our scheme. The $\left| e
\right\rangle \to \left| r \right\rangle $ transition is driven by a
classical laser pulse with coupling $\Omega $, and the $\left| r
\right\rangle \to \left| g \right\rangle $transition is driven by the
quantized cavity mode with coupling $g$. Both the classical laser pulse and
the cavity mode are detuned from their respective transitions by the same
amount $\Delta $.}
\label{fig1}
\end{figure}

In this paper, each of the atoms is a three-level system, which has
two ground states $\left| g \right\rangle ,\left| e \right\rangle
$(e.g. hyperfine ground states) and an excited state $\left| r
\right\rangle $ as depicted in Fig. \ref{fig1}. It is an adiabatic
evolution for the $\left| e \right\rangle \to \left| r \right\rangle
$transition, which is driven by a classical laser pulse with
coupling coefficient $\Omega $. The $\left| r \right\rangle \to
\left| g \right\rangle $transition is driven by the quantized cavity
mode with coupling coefficient $g$. Both the classical laser pulse
and the cavity mode are detuned from their respective transitions by
the same amount $\Delta $. Assuming the atom is trapped in a
specific position in the cavity, and the coupling coefficients
$\Omega $ and $g$ are constant during the interaction. In the case
of $\Omega g / \Delta ^2 < < 1$, the upper level $\left| r
\right\rangle $ can be decoupled from the evolution. When $\Delta >
> \gamma $ the spontaneous decay rate $\gamma $ from $\left| r
\right\rangle $ level can be neglected~\cite{Bose2}. In the
interaction picture and suppose $\Omega = g$, the effective
Hamiltonian of the atom interacted with the cavity is
\begin{equation} \label{eq1}
H_{eff} = i\delta (a\left| e \right\rangle \left\langle g \right| - a^ +
\left| g \right\rangle \left\langle e \right|) - ika^ + a,
\end{equation}
where $\delta = g\Omega / \Delta $, $a$ and $a^ + $ are the annihilation and
creation operators of the cavity mode, $k$ is the photon decay rate from the
cavity.

\begin{figure}[tbp]
\includegraphics[scale=0.4]{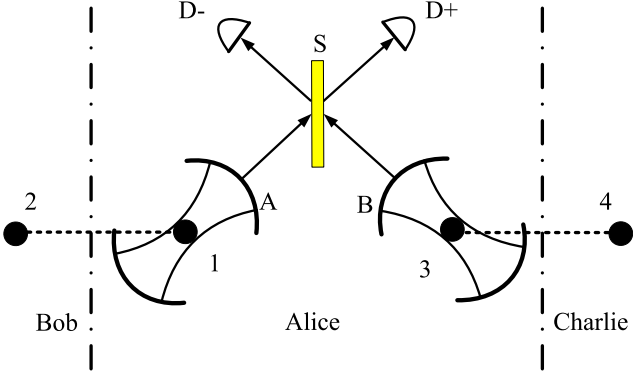}
\caption{The setup is adapted to concentrate two nonmaximally
entangled pairs into one maximally entangled. Atoms 1 and 3 are
trapped in the optical cavities $A$ and $B$, respectively. $S$ is a
50/50 beam splitter and $ \mbox{D}_\pm $ are single-photon
detectors. } \label{fig2}
\end{figure}

In QIP, maximally entangled atomic states are usually distributed among
users in different locations. Because of nonperfect generation scheme, the
entangled states of the distant atoms may be in the nonmaximally entangled
states. We can suppose that the entangled states of two pairs atoms
\begin{subequations}
\begin{equation}  \label{eq2a}
\left| \Phi \right\rangle _{12} = a\left| e \right\rangle _1 \left| g
\right\rangle _2 + b\left| g \right\rangle _1 \left| e \right\rangle _2 ,
\end{equation}
\begin{equation}  \label{eq2b}
\left| \Phi \right\rangle _{34} = c\left| e \right\rangle _3 \left| g
\right\rangle _4 + d\left| g \right\rangle _3 \left| e \right\rangle _4 ,
\end{equation}
\end{subequations}
where $a$, $b$,$c$ and $d$ are the normalization coefficients. To
extract maximally entangled states from the nonmaximally entangled
states, we can use the setup in Fig. \ref{fig2}. Atoms1 and 3 are
trapped in the optical cavities $A$ and $B$, respectively. Atoms 1
and 3, cavities $A$ and $B$, beam splitter $S$ and two detectors
belong to Alice, atom 2 belongs to Bob, and atom 4 belongs to
Charlie. The cavities are prepared in vacuum states initially.

Then, the initial states of system-$A$ (cavity $A$ and atoms 1, 2) and
system-$B$ (cavity $B$ and atoms 3, 4) are
\begin{subequations}
\begin{equation}  \label{eq3a}
\left| \Psi \right\rangle _{12A} = (a\left| e \right\rangle _1 \left| g
\right\rangle _2 + b\left| g \right\rangle _1 \left| e \right\rangle _2
)\left| 0 \right\rangle _A ,
\end{equation}
\begin{equation}  \label{eq3b}
\left| \Psi \right\rangle _{34B} = (c\left| e \right\rangle _3 \left| g
\right\rangle _4 + d\left| g \right\rangle _3 \left| e \right\rangle _4
)\left| 0 \right\rangle _B .
\end{equation}
\end{subequations}
Alice applies two same classical laser pulses on atoms 1 and 3,
respectively, to switch on the effective Hamiltonian $H_{eff} $ in
the System-$A$ and System-$B$ simultaneously. Choosing the
interaction time $t_1
$, which satisfies $\tan \frac{\Omega _k t_1 }{2} = - \frac{\Omega _k }{k}$%
(where $\Omega _k = \sqrt {4\delta^2 - k^2} )$, the states of system-$A$ and the
system-$B$ evolve into
\begin{subequations}
\begin{equation}  \label{eq4a}
\left| {\Psi }^{\prime}\right\rangle _{12A} = \frac{(a\alpha \left|
g \right\rangle _2 \left| 1 \right\rangle _A + b\left| e
\right\rangle _2 \left| 0 \right\rangle _A )\left| g \right\rangle
_1}{\sqrt {\vert a\vert ^2\alpha ^2 + \vert b\vert ^2}},
\end{equation}
\begin{equation}  \label{eq4b}
\left| {\Psi }^{\prime}\right\rangle _{34B} = \frac{(c\alpha \left|
g \right\rangle _4 \left| 1 \right\rangle _B + d\left| e
\right\rangle _4 \left| 0 \right\rangle _B )\left| g \right\rangle
_3}{\sqrt {\vert c\vert ^2\alpha ^2 + \vert d\vert ^2}},
\end{equation}
\end{subequations}
where $\alpha = - \frac{2\delta}{\Omega _k }e^{ - \frac{kt_1 }{2}}\sin \frac{%
\Omega _k t_1 }{2}$. The successful probability of evolvement are
$P_A = (\vert a\vert ^2\alpha ^2 + \vert b\vert ^2)$and $P_B =
(\vert c\vert ^2\alpha ^2 + \vert d\vert ^2)$, respectively. At the
time $t_1 $, the joint state of atoms 2, 4, and cavity $A$, $B$
becomes
\begin{widetext}
\begin{equation}
\label{eq5}\left| {\Psi \left( {t_1 } \right)} \right\rangle _{24AB}
= \frac{(a\alpha \left| g \right\rangle _2 \left| 1 \right\rangle _A
+ b\left| e \right\rangle _2 \left| 0 \right\rangle _A ) \times
(c\alpha \left| g \right\rangle _4 \left| 1 \right\rangle _B +
d\left| e \right\rangle _4 \left| 0 \right\rangle _B)}{\sqrt { P_A
P_B }},
\end{equation}
the success probability of this step is $P_{suc1} = P_A P_B $. If we select $%
\Omega _k > > k$, $P_{suc1} \approx 1$.

Now we consider the detection stage in which we make two single-photon
detectors [13, 14]. Alice will wait for one click at $\mbox{D}_{+}$ or $%
\mbox{D}_{-}$ for a time interval $t_{2}$, at a time $t_{j}$ in the
detection
stage ($t_{j}\leq t_{2})$, the joint state of atoms 2, 4, and cavities $A$, $%
B$ evolves into~\cite{plenio2}
\begin{equation}
\label{eq6} \left| {\Psi (t_j )} \right\rangle _{24AB} =
\frac{a\alpha e^{ - kt_{^j} }\left| g \right\rangle _2 \left| 1
\right\rangle _A + b\left| e \right\rangle _2 \left| 0 \right\rangle
_A }{\sqrt {\vert a\vert ^2\alpha ^2e^{ - 2kt_{^j} } + \vert b\vert
^2}}\times \frac{c\alpha e^{ - kt_j }\left| g \right\rangle _4
\left| 1 \right\rangle _B + d\left| e \right\rangle _4 \left| 0
\right\rangle _B}{\sqrt {\vert c\vert ^2\alpha ^2e^{ - 2kt_j } +
\vert d\vert ^2}}.
\end{equation}

If only one of the detectors $\mbox{D}_\pm $ clicks, it corresponds
to the action of the jump operators $(a_A \pm a_B ) / \sqrt 2$ on
the joint state $\left| {\Psi \left( {t_2 } \right)} \right\rangle
_{24AB} $, then the joint state of entire system becomes
\begin{equation}
\label{eq7} \left| {\Psi \left( {t_2 } \right)^\pm } \right\rangle
_{24AB} = \frac{ac\alpha e^{ - kt_2 }\left| g \right\rangle _2
\left| g \right\rangle _4 (\left| 0 \right\rangle _A \left| 1
\right\rangle _B \pm \left| 1 \right\rangle _A \left| 0
\right\rangle _B )+ (ad\left| g \right\rangle _2 \left| e
\right\rangle _4 \pm bc\left| e \right\rangle _2 \left| g
\right\rangle _4 )\left| 0 \right\rangle _A \left| 0 \right\rangle
_B}{\sqrt {\left| {ad} \right|^2 + \left| {bc} \right|^2 + 2\vert
ac\vert ^2\alpha ^2e^{ - 2kt_2 }}},
\end{equation}

If the $\mbox{D}_ - $ clicked, Alice lets Charlie give $\left| g
\right\rangle _4 $ an extra phase shift $\pi$ with respect to
$\left| e \right\rangle _4 $ by classical communication channel; if
the $\mbox{D}_ + $ clicked, no extra phase shift is required.

If the initial states of atoms 1, 2 and 3, 4 satisfy the condition
$a = c$, $b = d$, the state $\left| {\Psi \left( {t_2 } \right)^ + }
\right\rangle _{24AB} $ will become
\begin{equation}
\left\vert \Psi \left( t_{2}\right) ^{+}\right\rangle _{24AB}=\frac{%
a^{2}\alpha e^{-kt_{2}}\left\vert g\right\rangle _{2}\left\vert
g\right\rangle _{4}\frac{1}{\sqrt{2}}\left( \left\vert
0\right\rangle _{A}\left\vert 1\right\rangle _{B}+\left\vert
1\right\rangle _{A}\left\vert 0\right\rangle _{B}\right)
+ab\frac{1}{\sqrt{2}}\left( \left\vert g\right\rangle _{2}\left\vert
e\right\rangle _{4}+\left\vert e\right\rangle _{2}\left\vert
g\right\rangle _{4}\right) \left\vert 0\right\rangle _{A}\left\vert
0\right\rangle _{B}}{\sqrt{\left\vert ab\right\vert ^{2}+\left\vert
a\right\vert ^{4}\alpha ^{2}e^{-2kt_{2}}}}  \label{eq8}
\end{equation}
\end{widetext}

Now, we consider the state $\left| {\Psi \left( {t_2 } \right)^ + }
\right\rangle _{24AB} $ by tracing over the cavities $A$ and $B$,
the reduced density matrix of atoms 2 and 4 is
\begin{equation}
\rho _{24}=\frac{\left\vert ab\right\vert ^{2}\left\vert \Phi \right\rangle
_{2424}\left\langle \Phi \right\vert +\left\vert a\right\vert ^{4}\alpha
^{2}e^{-2kt_{2}}\left\vert g\right\rangle _{2}\left\vert g\right\rangle
_{44}\left\langle g\right\vert _{2}\left\langle g\right\vert }{\left\vert
ab\right\vert ^{2}+\left\vert a\right\vert ^{4}\alpha ^{2}e^{-2kt_{2}}}
\label{eq9}
\end{equation}%
where $\left\vert \Phi \right\rangle
_{24}=\frac{1}{\sqrt{2}}(\left\vert g\right\rangle _{2}\left\vert
e\right\rangle _{4}+\left\vert e\right\rangle _{2}\left\vert
g\right\rangle _{4})$, which is the ultimate result of
concentration. The total successful probability of obtaining the
entangled state in Eq. (\ref{eq9}) is $P_{success}=(\left\vert
{ab}\right\vert ^{2}+\left\vert a\right\vert ^{4}\alpha
^{2}e^{-2kt_{2}})\alpha
^{2}e^{-2kt_{2}}(1-e^{-2kt_{2}})$, and the fidelity of the obtained state $%
\left\vert \Phi \right\rangle _{24}$ is $F=\left\vert b\right\vert
^{2}/(\left\vert b\right\vert ^{2}+\left\vert a\right\vert
^{2}\alpha ^{2}e^{-2kt_{2}})$. In the case of $\left\vert
{a/b}\right\vert <<1$, the fidelity will be $F\approx 1$.

In conclusion, we have proposed an entanglement concentration scheme for
unknown atomic entangled states with a finite probability via cavity decay.
Compared with other schemes, our scheme has the following advantages: (a) In
our scheme, the atomic state is used as stationary qubit and photonic state
as flying qubit, which is more feasible experimentally. thus the distant
entangled states concentration via high quality fiber will be convenient.
(b) Taking cavity decay into consideration, it is more practical to discuss
entanglement concentration.

\begin{acknowledgements}
This work is supported by the Natural Science Foundation of the Education
Department of Anhui Province under Grant No: 2004kj005zd and Anhui
Provincial Natural Science Foundation under Grant No: 03042401 and the
Talent Foundation of Anhui University.
\end{acknowledgements}

\end{document}